# SEMANTIC WEB TECHNIQUES FOR YELLOW PAGE SERVICE PROVIDERS

Raghu Anantharangachar[1] and Ramani Srinivasan[2]

[1]Hewlett Packard Laboratories, India
[1]International Institute of Information Technology, Bangalore
araghu@hp.com
[2]International Institute of Information Technology, Bangalore
ramani.srini@gmail.com

## ABSTRACT

*Applications providing "yellow pages information" for use over the web should ideally be based on structured information. Use of web pages providing unstructured information poses variety of problems to the user, such as use of arbitrary formats, unsuitability for machine processing and likely incompleteness of information. Structured data alleviates these problems but we require more. Capturing the semantics of a domain in the form of an ontology is necessary to ensure that unforeseen application can easily be created at a later date. Very often yellow page systems are implemented using a centralized database. In some cases, human intermediaries accessible over the phone network examine a centralized database and use their reasoning ability to deal with the user's need for information. Centralized operation and considerable central administration make these systems expensive to operate. Scaling up such systems is difficult. They behave like isolated systems and it is common for such systems to be highly domain specific, for instance systems dealing with accommodation and travel. This paper explores an alternative – a highly distributed system design meeting a variety of needs – considerably reducing efforts required at a central organization, enabling large numbers of vendors to enter information about their own products and services, enabling end-users to contribute information such as their own ratings, using an ontology to describe each domain of application in a flexible manner for uses foreseen and unforeseen, enabling distributed search and mash-ups, use of vendor independent standards, using reasoning to find the best matches to a given query, geo-spatial reasoning and a simple, interactive, mobile application/interface. We view this design as one in which vendors and end-users do the bulk of the work in building large distributed collections of information in a Web 2.0 style. We give importance to geo-spatial information and mobile applications because of the very wide-spread use of mobile phones and their inherent ability to provide some information about the current location of the user. We have created a prototype using the Jena Toolkit and geo-spatial extensions to SPARQL. We use simple and shallow reasoning to give inferred information in addition to explicitly stored information. We have tested this prototype by asking a group of typical users to use it and to provide structured feedback. We have summarized this feedback in the paper. We believe that the technology can be applied in many contexts in addition to yellow page systems. The essential features are the involvement of a set of creators of information and a set of end-users, the two sets not being mutually exclusive and the use of an ontology that can span a number of domains in the context of product and service vendors.*

## KEYWORDS

*Semantic web, Triples, Resource Description Framework, Web Service*

## 1. BACKGROUND

Many of us consult the web for obtaining answers to our questions on a regular basis. Usually, we use search engines directly to find answers to our questions, and sometimes we access call centers that provide local information over the phone. Though we might be clear in our mind about what we are looking for or searching for, we find it often difficult to communicate a search request clearly. Even if the question is phrased carefully, the response is not necessarily precise. Systems designed to support "browsing" are significantly different from those designed to get precise answers. Most search engines primarily support browsing, and are not designed to give precise and relevant answers to every question. Whenever we search the web, we get numerous results that are ranked in some order and displayed to us; this is a bit like being given a set of reading materials in response to our query. The rest is up to the individual user.

Database researchers have used structured information to improve the quality of the responses to their query. However, there are a variety of practical problems in networking a set of distributed, heterogeneous databases over the internet owned and operated by a large number of different organizations. In addition, the database users use the web as a pipe for carrying information. Semantic web brings a new vision of having web scale information repositories.

## 2. INTRODUCTION

Consider a scenario, wherein we are looking for a hotel which provides lunch in Adugodi, a locality in South Bangalore, in the price range below Rs.100. It is difficult to find a suitable answer.

As we can see from the output, the search provides results that are not directly related to what we are looking for. The suggested results are related to property and not hotels. In other words, the results are not relevant to the query.

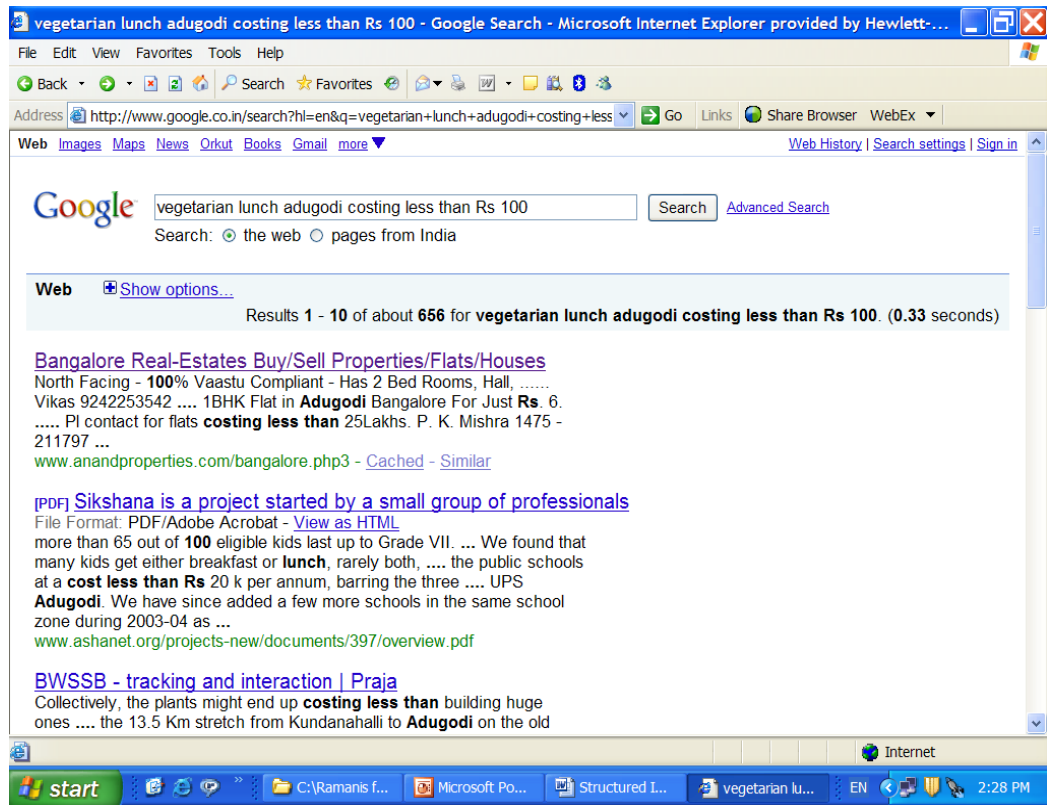

**Figure 1: Google query output**

We tried the query with the justdial.com website, operated by a service provider who offers yellow-page style local information about Bangalore. They have a list of customers who have registered with them, and give their contact information to people who call the "audio search engine" (as they have named themselves). We got responses which did not provide us the answer to our question. We were actually looking for restaurants, but the "answers" are not at all relevant to this question.

See figure below:

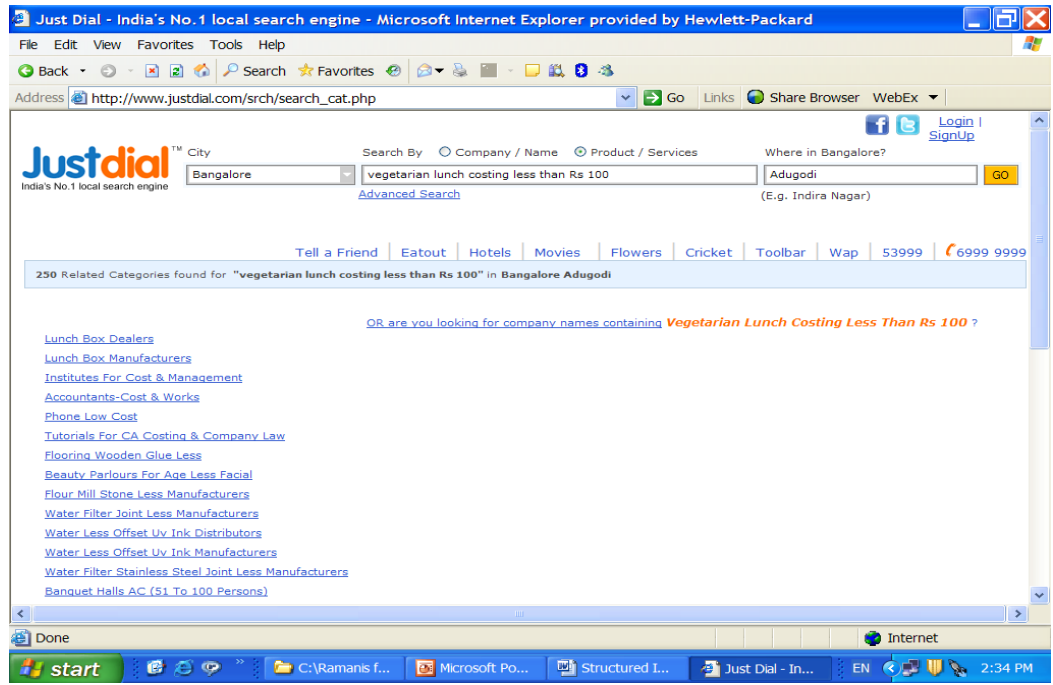

**Figure 2: Justdial query output**

We then went ahead and tried the same query with Google Maps. We wanted to know if Google Maps provides more accurate information to us, which can help us answer our query.

In this case, we again got irrelevant information. The suggested restaurants are at least thirty minutes by car from the place where we were looking for a restaurant, and a typical lunch at the restaurant suggested first costs three times the limit we specified. In addition, it is not a vegetarian restaurant. This may be due to the fact that the maps application provides usable information on only those restaurants which register with it. This is not what the television viewers have come to believe - that the web is full of riches to be tapped at the touch of a button – all created through a Web 2.0 mechanism!

The output is given in the figure.

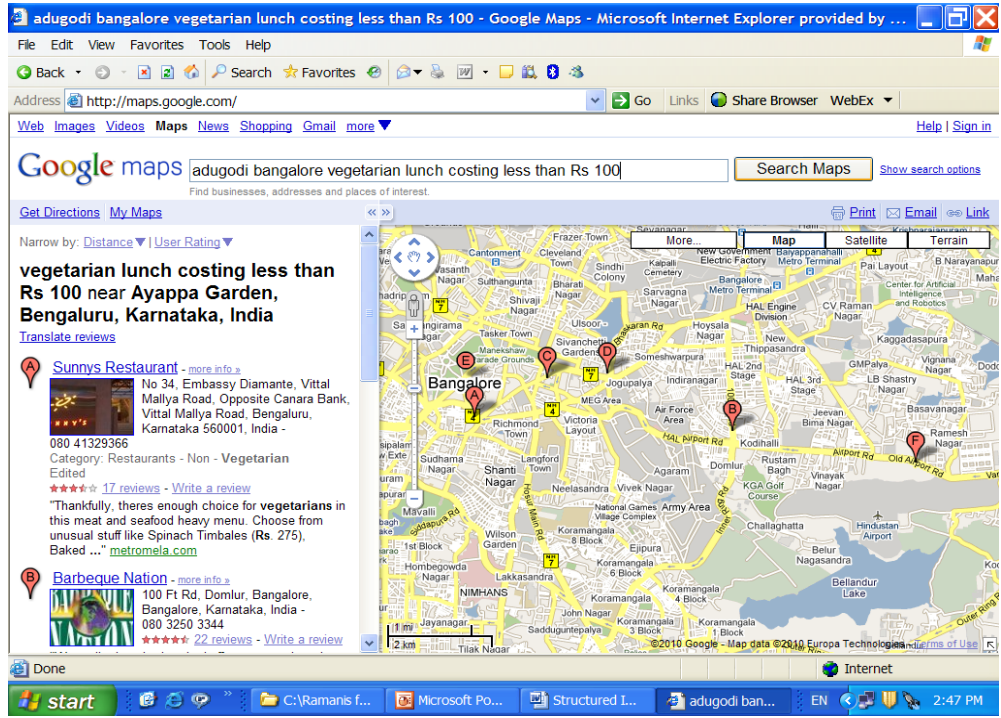

However, maps applications like the one mentioned above show user created tags which are sometimes misleading. It is not uncommon to find that a restaurant or hotel being shown to be at a place several kilometres away from the real address.

## 3. BROWSING VERSUS QUESTION ANSWERING

The web is a huge repository of information, and we argue that it is difficult for the common man to find suitable answers to his questions using a browser. This is due to many reasons, the important ones are given below:

### *Tool Limitation*

As mentioned earlier, the browser & search engine combination is a tool that has been created to enable the users to browse the web. It is not designed to give precise and reliable answers every time. Whenever we search the web, we get numerous results that are ranked in some order and displayed to us; this is a bit like being shown a relevant rack in the library. The rest is up to the individual user. Extracting the relevant fragment from the web page requires manual scrutiny. In addition, extracting the exact answer from the fragment poses its own problem.

*Problem*

Search is a complex process due to the following factors:

- The ability to use the relations in the search query is not possible using a browser. The user has several relations between the keywords in mind while typing in the keywords, however, existing search engines like Google do not support relations in the query. This results in irrelevant results.
- The relevance of the "hits" is low, partly because the user's questions are not always unambiguous.
- Unsatisfactory "hits" can distract the searcher leading him to read other possibly interesting information unrelated to his original search.
- The users need to extract relevant fragments of information from each web site, and synthesize answers to their questions. This involves locating the websites, understanding their content, and finding answers. This puts a significant cognitive load on the user.
- The search by the user takes a fair amount of time, as it involves manual processing
- Information might be scattered on the website at different places
- The user may not get to the relevant page at all, though he might be getting a large number of pages for his query.

*Mobile Connectivity*

If the user is using a mobile device, there are additional limitations and restrictions, as the mobile devices get frequently disconnected from a session. Mobile phones are not suitable for reading several pages of text to answer one question. All these factors further downgrade the user experience in obtaining good answers, and increase the time taken for getting information.

CRITERIA FOR A GOOD QUESTION ANSWERING SYSTEM

a) Responses should be in real time, and not require expertise in query design or significant browsing.

b) Imprecise, incorrect answers are worse than no answers. The user should not be given misleading or contradictory information[25].

c) The system should provide a high degree of recall as well as precision; for this purpose the interface should provide for the creation of ideally unambiguous queries.

d) Answers distributed across multiple sources require fusion techniques that combine partial answers from different sources

e) The answer should be relevant within a specific context. Context can be used to clarify a question and resolve ambiguities.

f) Question answering should be personalized to the user using information about him.

**KEY CONTRIBUTIONS OF OUR PAPER**

We propose a simple process oriented Yellow Page Service Provider system that uses semantics of query to retrieve precise results for the user. We argue that by using semantics in the query, and by storing the data in a triple format, we are able to obtain very precise answers to user's queries. We are also working on converting the natural language text into semantic triple format automatically, which we describe in another paper.

## YELLOW PAGES SERVICE PROVIDERS

The Yellow Page Service providers (YPSPs) collect the information about various business establishments and offer that information through a web portal, typically for a price. They encourage business establishments to register themselves, and provide information about their business which is relevant to end users. Subsequently, this information gets exposed to end users through a variety of interfaces - the important ones being a web portal or a telephonic channel. The information is stored in a persistent store, and accessed by the yellow pages service providers to answer the queries posed by their end users. The business model of such a typical Yellow Pages Service Provider is given.

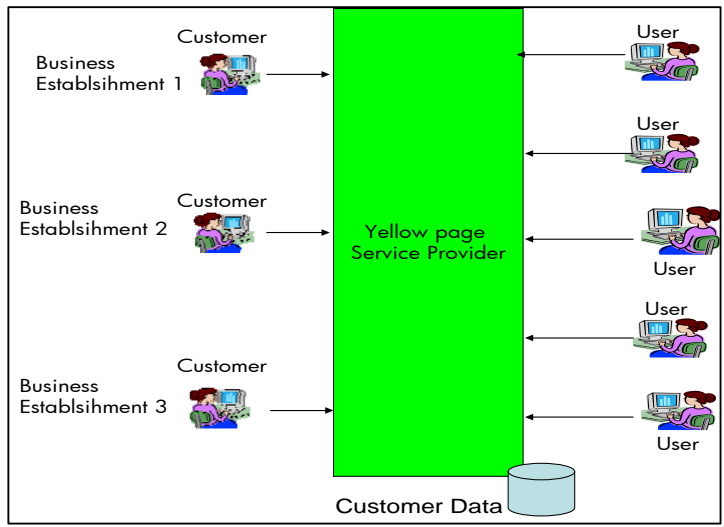

As we can see from the above diagram, there are two types of activities here. The first part is the registration/update phase wherein the business establishments register themselves with the YPSP and provide information about their business. The YPSP stores all this information. Then, the second part of the business is to enable end users to search for specific information and use the information provided by the business establishments to answer their queries.

Most YPSPs charge a fee from their customers (who typically represent various business establishments) and provide a service to the users for free. By providing their information to the YPSPs, various business establishments are able to increase their reach to end users, and thereby increase their customer base. It is similar to an advertisement based model, wherein the business establishments pay a fee to get their particulars included in the answer given to the user.

## RESEARCH PROBLEMS ADDRESSED IN THIS PAPER

1. How do we simplify information retrieval, especially for naïve users with specific questions?

2. How do we help naïve users to disambiguate their queries and articulate them in a way that a system can process them without hit and miss?

3. How do we create and maintain a distributed large scale system with tens of thousands of information providers?

4. How do we do all this at a low enough cost to enable every shop to be on the system, and to support users on millions of mobile phones?

## RELATED WORK

The related work seems to be from three major streams. These are given below:

### *Databases*

Database researchers have tried to use databases to work with a question answering interfaces. In this context, the database is used to store information and database queries are created by the interface to retrieve relevant answers from the databases. The user typically communicates through the interface in natural language.

Researchers in the database community have tried using natural language interfaces to the databases[7], [9], [10], [13], and as well migration of data contained in the databases to the RDF triple format to interface with the semantic web. In fact, as early as 1993[1], an effort was underway to identify the problems in providing a natural language interface to databases which highlighted application related issues, database related issues, interaction related issues and the language related issues. Notable efforts in this stream include Baseball, Lunar (1960), MASQUE SQL (1993) and PRECISE (2003). DLDB[2] is an effort in this direction that examines options to reformat the relational data existing in relational database to a semantic format. In this effort, an attempt is made to build a knowledgebase that can perform DAML-IOL inference, especially from the point of view of various database schemas to store RDF data.

### **Open Domain Question Answering**

This stream deals with web based question answering considering the entire web as the corpus. This stream uses many information retrieval techniques to increase the relevance of the results. Some of the important efforts in this stream include ACQUAINT (2003), LASSO, Askjeeves[3], EasyAsk[4] and AnswerBus[5]. Some of them provide web links as answers and some others provide documents (ex. askjeeves). These approaches are based on building a corpus of questions and answers and using the natural language query to identify appropriate answers. In case the answer is readily not available, the question is posted back to a forum, and once one of the forum members has answered the question, the answer is provided back to the user, and added to the corpus. The researchers have used a number of information retrieval based methods for question answering[16], and as well information fusion[14] for question answering.

### *YPSPs and Similar Service Providers*

There are many service providers who offer information over phone and web to their users. These include providers like justdial.com, asklaila.com, askjeeves.com and the like. These providers store the information about their customers (who are registered with them), and make this information available to end users over phone and web. An effort like answers.com is another related work, which serves like a bulletin board containing a collection of questions and answers.

### *Semantic web*

Semantic web provides a good technology for creating question answering systems; using semantics increases the chances of retrieving an appropriate answer for a particular question. The semantic web stream essentially deals with using concepts involving RDF and OWL representations, and inferences to provide meaningful responses to user queries. This is a very active stream, and there are many research projects and prototypes which are attempting to solve problems similar to ours. The major efforts in this stream include START[26], DIMAP, PiQaSSo[19], ORAKEL[23] and XeOML[22]. Lopez et al[27] provide a detailed survey of the various efforts in the semantic question answering.

An important effort in this direction is the "Talking to the Semantic Web" project[12], in which the researchers studied the suitability of various interfaces drawn from various other projects for casual users. This effort includes Querix[20], NLP-Reduce[15], Ginseng[11] and Semantic Crystal. NLP-Reduce is another effort that attempts to use a natural language interface and generates SPARQL query statements, and returns matched output. Querix is a system that interacts with the user in case of ambiguities, and initiates a dialog to obtain appropriate inputs in natural language format, and outputs relevant answers.

The other major efforts include Nalix[3] and PANTO[18]. Ginseng[11] is a system that supports natural language query over XML databases. PANTO is an initiative that accepts queries in natural language format, and outputs SPARQL queries.

Yahoo answers[24] is an asynchronous question answering system, wherein experts actually provide answers to the questions. These questions are received over the web, and it is likely that some questions are not at all answered. Aqualog[17] is an important effort in the direction of creating a question answering system using semantic web ontologies. IRSAW[21] is a system that provides annotation of documents for question answering.

## HOTEL ONTOLOGY

We have created a Hotel ontology by using some of the standard attributes that are used to describe a hotel using protégé. A sample class in the ontology is given below:

```
<?xml version="1.0"?>
<rdf:RDF
    xmlns:p2="http://www.owl-ontologies.com/restaurant#"
    xmlns:rdf="http://www.w3.org/1999/02/22-rdf-syntax-ns#"
    xmlns:rest="http://localhost/restaurant.owl#"
    xmlns:p1="http://www.owl-ontologies.com/assert.owl#"
    xmlns:owl="http://www.w3.org/2002/07/owl#"
    xmlns:xsd="http://www.w3.org/2001/XMLSchema#"
    xmlns:rdfs="http://www.w3.org/2000/01/rdf-schema#"
    xmlns="http://www.owl-ontologies.com/Restaurant.owl#"
  xml:base="http://www.owl-ontologies.com/Restaurant.owl">
  <owl:Ontology rdf:about=""/>
  <owl:Class rdf:ID="Restaurant"/>
  <owl:DatatypeProperty rdf:ID="hasLocation">
    <rdfs:range rdf:resource="http://www.w3.org/2001/XMLSchema#string"/>
    <rdfs:domain rdf:resource="#Restaurant"/>
  </owl:DatatypeProperty>
  <owl:DatatypeProperty rdf:ID="hasName">
```

```xml
    <rdfs:domain rdf:resource="#Restaurant"/>
    <rdfs:range rdf:resource="http://www.w3.org/2001/XMLSchema#string"/>
  </owl:DatatypeProperty>
  <owl:DatatypeProperty rdf:ID="hasCost">
    <rdfs:domain rdf:resource="#Restaurant"/>
    <rdfs:range rdf:resource="http://www.w3.org/2001/XMLSchema#int"/>
  </owl:DatatypeProperty>
  <owl:DatatypeProperty rdf:ID="hasMealType">
    <rdfs:range rdf:resource="http://www.w3.org/2001/XMLSchema#string"/>
    <rdfs:domain rdf:resource="#Restaurant"/>
  </owl:DatatypeProperty>
  <owl:FunctionalProperty rdf:ID="hasFoodType">
    <rdfs:domain rdf:resource="#Restaurant"/>
    <rdfs:range rdf:resource="http://www.w3.org/2001/XMLSchema#string"/>
    <rdf:type rdf:resource="http://www.w3.org/2002/07/owl#DatatypeProperty"/>
  </owl:FunctionalProperty>
  <Restaurant rdf:ID="Darshini">
    <hasFoodType rdf:datatype="http://www.w3.org/2001/XMLSchema#string"
    >Veg</hasFoodType>
    <hasName xml:lang="en">Darshini</hasName>
    <hasLocation xml:lang="en">Koramangala</hasLocation>
    <hasLocation xml:lang="en">Vijayanagar</hasLocation>
    <hasMealType xml:lang="en">Lunch</hasMealType>
    <hasCost rdf:datatype="http://www.w3.org/2001/XMLSchema#int"
    >100</hasCost>
  </Restaurant>
  <Restaurant rdf:ID="Kamat">
    <hasMealType xml:lang="en">Dinner</hasMealType>
    <hasName rdf:datatype="http://www.w3.org/2001/XMLSchema#string"
    >Kamat</hasName>
    <hasLocation xml:lang="en">Gandhinagar</hasLocation>
    <hasCost rdf:datatype="http://www.w3.org/2001/XMLSchema#int"
    >50</hasCost>
    <hasFoodType rdf:datatype="http://www.w3.org/2001/XMLSchema#string"
    >Veg</hasFoodType>
  </Restaurant>
```

```xml
<owl:AllDifferent>
  <owl:distinctMembers rdf:parseType="Collection">
    <Restaurant rdf:about="#Kamat"/>
  </owl:distinctMembers>
</owl:AllDifferent>
<Restaurant rdf:ID="Upahar">
  <hasLocation rdf:datatype="http://www.w3.org/2001/XMLSchema#string"
  >Vijayanagar</hasLocation>
  <hasFoodType rdf:datatype="http://www.w3.org/2001/XMLSchema#string"
  >Veg</hasFoodType>
  <hasMealType xml:lang="en">Breakfast</hasMealType>
  <hasCost rdf:datatype="http://www.w3.org/2001/XMLSchema#int"
  >25</hasCost>
  <hasName xml:lang="en">Upahar</hasName>
</Restaurant>
</rdf:RDF>
```

We have retained a few instances of the classes to show the completeness of the ontology. As we can see there are several datatype properties that are used to capture the various attributes of a hotel. We focus on Hotel class in this paper, as it helps us to explain our approach with respect to a single class. However, the logic explained in this paper can be used with reference to any other class as well (say Hospital).

## OUR SOLUTION

As we saw in the earlier chapter on Business Models, there are two types of activities. The first is about the registration/update of business data and the second step is about end user search. Our approach for creating a solution addresses both these steps.

### 1.  *Customer Registration Phase:*

Customer Registration phase involves various business establishments communicating with the YPSP's system over the web and registering with the YPSP by providing information structured by the interface provided by the YPSP.

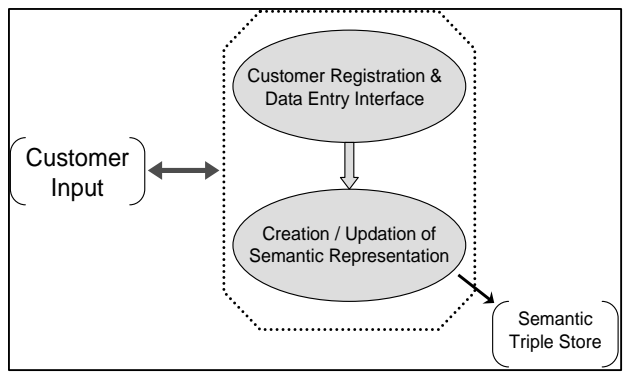

The architecture for the registration is given in the following fig.

The customer or the customer representative performs the following tasks to provide information about their respective offerings:

1. The customer initially connects to the website of the YPSP and logs into the system using his username and password.

2. The customer is then taken to a simple user interface which allows him to provide information about his business. He starts with selecting the business type (which could be Hospital, College, Restaurant or any other, listed by the YPSP). Based on the selected business type subsequent menus are customized to show the fields that are relevant to that particular business. For example, in the case of Restaurants the customer is asked to enter information about the type of food provided, whereas in the case of Colleges the customer is asked to enter information about the educational programs offered, their pre-requisites, duration etc. Once all the inputs are provided by the customer, the customer presses the submit button, and the information is submitted to the YPSP server.

3. The YPSP server now converts the information provided by the customer into semantic representations, and stores the information as a set of triples in a persistent store.

4. Once this conversion and the subsequent update into the triple store is complete, the customer data is available for the end users to search. At this point, the customer is notified about the successful completion of the registration process.

### *END USER INITIATED SEARCH PHASE:*

This is the phase in which the various end users perform web based querying using an interactive user interface, and retrieve responses from the YPSP.

The solution architecture to address the end user initiated search is given here:

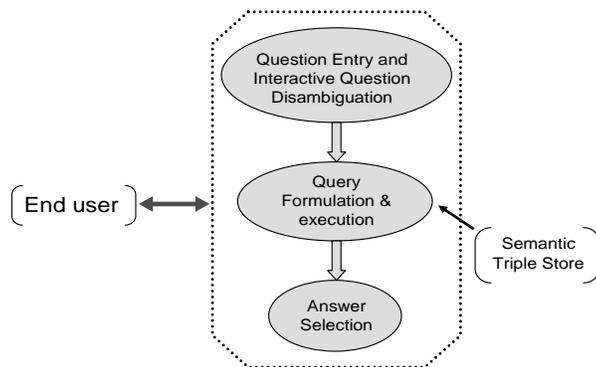

Our approach includes the following steps:

- Create a simple user interface to help the naïve user to phrase his query. This browser based interface does not require any special client application to be installed. This interface is designed to enable the user to enter his inputs in a structured manner, take the user's inputs and display responses at each step. The user stops the interaction when he thinks the query inputs lead to good enough results.

- The interface helps the user to build his query by asking him for specific information during various steps in the above mentioned activity. For instance, one of the first steps would be to ask the user to choose a specific business type/entity (and thus automatically select an appropriate ontology/schema based on this input). If the user is looking for a hotel, he selects hotel. Once the user has selected hotel, we use this as the key entity, and retrieve its attributes from the semantic representation. We then ask the user to provide his inputs in accordance with the ontology for hotels. The user selects the attributes to which he wants to provide values, and provides the values.

- Once the user has provided all the mandatory inputs and has pressed the submit button, the interface passes this information on to a back-end question answering engine, mapping the question to a semantic query. This mapping is done using simple logic - shallow reasoning where appropriate. The semantic query is created in SPARQL. By using the fields that are relevant to the ontology, we are able to generate precise queries in SPARQL which can return precise answers. In order to generate valid SPQRQL queries, we use a SPARQL query template at the server end, and ensure that the various form elements (displayed in the browser at the client end) are basically part of the ontology. In other words, the ontology elements form the shared information between the client and the server, and hence it is possible to use a SPARQL query template and generate valid queries.

- The semantic query engine then executes the semantic query, and gathers the results. The corpus for this query is the triple stores that are created for each website. It is possible that these triple stores are hosted at different sites. We need to take care of routing the query to the semantic engine that contains the most relevant semantic store for that particular ontology/schema.

- The results of the query are synthesized and provided to the user. The result displayed to the user through a simple user interface enables the user to continue his querying if he is not satisfied with the results. For example, once the user has looked at the results, he might decide that he wants to provide additional input to make his query less ambiguous. We provide an option for the user to go back to the query window, and refine his query or provide additional optional inputs. Then, once the user presses the submit button, we process the query, and provide him with the results.

- The corpus used for consultation should be created using a Web2.0 process but the customers creating the data pay for it and take the responsibility for the content. All end users can query and registered end users can also add comments and reviews.

- The corpus is created as a distributed semantic representation (in the form semantic triples). For instance, a hotelier's association may host a triple store for hotels, allowing hotels to frequently update their entries. The information available to the system may be in the form of a few hundred triple stores, the hotelier's one being only one of them. Airlines and department stores could host their own triple stores, containing lots of information. We have created a tool and made it web enabled, so that the various website owners using a simple interface and a set of ontologies can create a semantic representation for their own offerings. In addition, we also support roles in the tool – to enable administrators to create additional attributes /relationships as and when required, and for the users to query the semantic stores.

- By using additional information that is available over the web (like the geographical information), it is possible to provide answers to queries that are otherwise difficult to answer. For example, when a user searches for a restaurant in Adugodi, it is possible that there are no restaurants in Adugodi. In that case, we should be able to provide a restaurant from Koramangala, which is adjacent to Adugodi geographically. This is possible as we use geographical information in our ontology. This is one example of the type of role played by reasoning. Another example is to offer the user information on similar establishments, when the ideal establishment indicated by his query does not exist.

**USER INTERFACE DESIGN**

A simple user interface for the user to select his choices is created keeping the following requirements in mind:

- The user interface shall be intuitive

- The user interface shall use choices from an ontology so that it ensures the use of standardized, unambiguous terms.
- Parts of the ontology for use with human computer interface should be domain specific
- The user interface shall be interactive and enable the user to keep his session active till he is satisfied with the answers.

Illustrations of the user interface is given below:

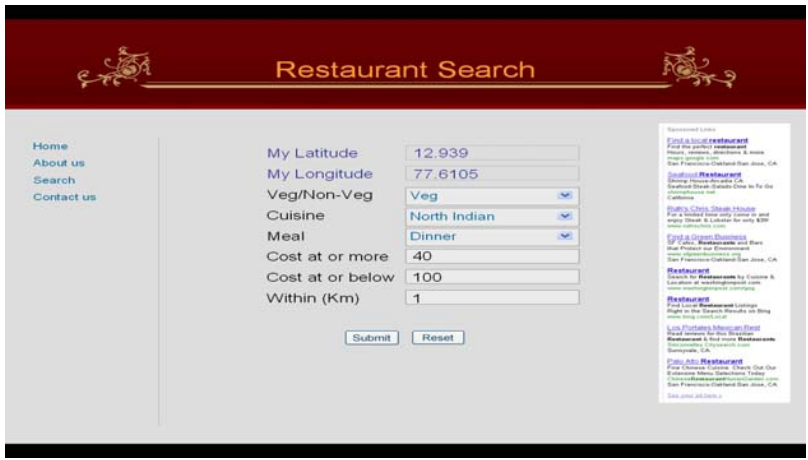

The response to this question is shown below:

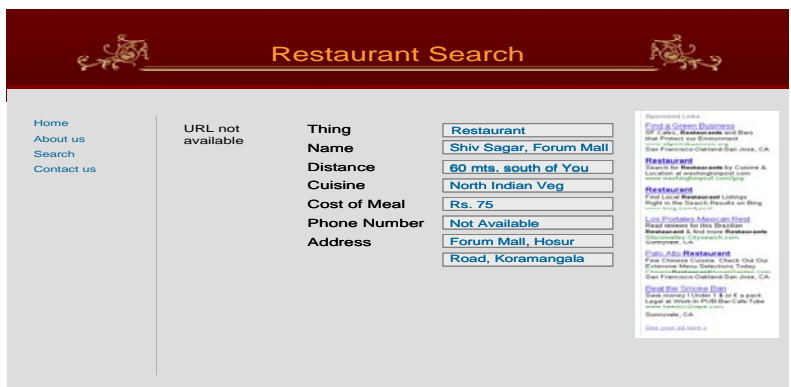

## SYSTEM ARCHITECTURE

In our system, we use a simple client and server model to create a service that is deployed in the cloud. Our solution consists of a client that is thin client that can run on any device as it is a simple HTML page. This thin client has javascript logic to read the various elements of the form, and convert them into a HTTP form. Please see the architectural diagram below:

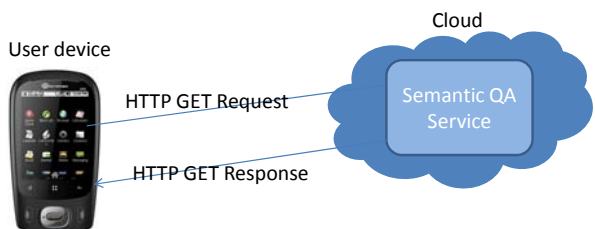

The Semantic QA Service receives the HTTP request from the client device, and then converts the HTTP request into SPARQL format. Once the user query is mapped to an appropriate

SPARQL query, then it invokes the jena toolkit to run the query on the semantic triple store. Once the execution of the query is complete, the results are processed, and returned to the user, as a HTTP GET response. A sample SPARQL query that is automatically generated is given below:

```
 1 PREFIX  geo:  <http://www.w3.org/2003/01/geo/wgs84_pos#>
 2 PREFIX  rest: <http://localhost:8080/>
 3 PREFIX  ext:  <java:org.geospatialweb.arqext.>
 4
 5 SELECT  ?name ?address
 6 WHERE
 7   { ?restaurant  ext:nearby        _:b0 .
 8     _:b0         <http://www.w3.org/1999/02/22-rdf-syntax-ns#first>  12.938147 ;
 9                  <http://www.w3.org/1999/02/22-rdf-syntax-ns#rest>   _:b1 .
10     _:b1         <http://www.w3.org/1999/02/22-rdf-syntax-ns#first>  77.609825 ;
11                  <http://www.w3.org/1999/02/22-rdf-syntax-ns#rest>   _:b2 .
12     _:b2         <http://www.w3.org/1999/02/22-rdf-syntax-ns#first>  5 ;
13                  <http://www.w3.org/1999/02/22-rdf-syntax-ns#rest>   <http://www.w3.org/1999/02/22-rdf-syntax-ns#nil> .
14     ?restaurant  rest:foodtype     "Veg" ;
15                  rest:mealtype     "Lunch" ;
16                  rest:address      ?address ;
17                  rest:name         ?name .
18   }
```

## HIGHLIGHTS OF OUR SYSTEM

1. Geospatial Reasoning: People often are interested in finding out relevant information while on the move, using their mobile phones. In these cases, it becomes difficult to expect them to enter a lot of information. So, our solution uses location information coming from their mobile phone, and helps us to provide an answer which is geographically close to the end users current position. In this case, the user only needs to enter what he is looking for (like restaurant), and the system will provide the user with an appropriate set of restaurants, and their addresses. He also has the option of identifying a neighbourhood elsewhere for the search.

   We use the GPS (Global Positioning Satellite) information to locate the establishment with increased precision. Our ontology contains geo-location details about each entity and we create an index based on the location coordinates which is used by the Google Geospatial extensions to SPARQL APIs [29]. These APIs help us to locate an establishment which is near the landmark specified in the query (or to the user's current geospatial coordinates). The list of restaurants thus obtained is later matched against the other constraints provided by the user. Once the information retrieval and the inference are complete, the list is provided to the user.
   We use our ontology as well as information provided by business establishments in creating appropriate responses to end users. Limited reasoning is used to ensure satisfactory responses.

2. Penalty Based Relaxation of Constraints: Many times, the search query posed by the user may not result in any answer – in which case, it makes sense to relax some of the constraints that were posed by the user, and arrive at an answer. This helps to get an answer that is close to what the user is looking for. For example, the user may be looking for a restaurant below price of Rs.100, and we may not find any result. By relaxing this condition slightly, it may be possible to obtain an answer. We perform constraint relaxation using rule based reasoning to find bestbet answer. However, every time we relax a constraint specified by the user, we associate a penalty to the result obtained. At the end, we present the result with the lowest penalty to the user. This ensures that the user gets the best answer. We can optionally provide the entire list to the user as well – sorted in the increasing order of penalty.

3. SPARQL End point support[28]: The support for only the user interface may not be sufficient, and by enabling a query over a web service it would be possible to build mashups over data that is geographically distributed, tapping by multiple semantic engines to do a part of the work each. Keeping this in mind, we support a SPARQL end point over which SPARQL queries can be sent to our engine to elicit appropriate responses. These SPARQL queries are sent over HTTP, and the responses are returned over HTTP. It is possible to develop new mashups using this facility. In addition, it is also possible to federate the semantic representation from our system along with the information coming from other SPARQL engines, and build mashups. For example, it is possible to combine local information with information coming from other sources (like airlines and so on) to provide the end user a greater variety of information.

## EXPERIMENTAL RESULTS

We have created a prototype to validate our proposal. Our prototype consists of a simple user interface which loads from a web server, an Apache Tomcat server in our prototype. A servlet created and deployed on the Apache Tomcat server invokes Jena APIs to search an ontology file (OWL file), and returns the result to the user. We created a small ontology consisting of instances of colleges, hospitals and restaurants. Subsequently, we used our prototype to query stored information, and asked a set of volunteers to search for information using any search engine and using our search interface. Subsequently, we asked them to fill up a questionnaire to help us evaluate our approach.

Questions Asked in the User Study:

The questions listed below are mostly pertinent to Bangalore. Adugodi, Koramangala, Jayanagar are all places within Bangalore.

1. Please find a restaurant in Adugodi which provides vegetarian lunch at Rs50-100 which is within 1 km of Forum Mall
2. Please find a restaurant in Jayanagar that provides a vegetarian Lunch for Rs.50 near main bus terminus
2. Please find a Nursery school in Jayanagar $3^{rd}$ block near Kadambam restaurant
3. Please find a hospital that treats accident victims in Rajajinagar $1^{st}$ block
4. Please find a hotel in Mumbai which is within 1 km of HP Office, which charges less than Rs.5000 per day.
5. Was 22 March 2010 a holiday in Bangalore?
6. Can I get a plane ticket to go from Bangalore to Mumbai tomorrow to reach there before 9AM non-stop?
7. Please find the Telephone number of an electrician in Vijayanagar, Bangalore who knows how to repair borewell pumps.
8. Where can I find a book shop that sells 9th standard textbooks (in Kannada) in Vijayanagar?
9. Please find a coconut plucker in Vijayanager
10. Look for a DVD Library which sells English blockbuster VCDs @ Rs.100 each or less in Vijayanagar
10. Please find a Doctor on demand - for visit to house for an eye checkup at home

Our findings are given below:

| SNo | Parameter | Our Search | Search Engines |
|---|---|---|---|
| 1 | Ease of use | High | Low |
| 2 | Typical response time | 5 seconds | 10 mins(approx) |
| 3 | Search precision | High | Low/Medium (many times search did not converge) |
| **4** | Query ambiguity | Low | Medium/High |

Some of our participants were not able to get answers to several queries without using our system, and some took a long time. In addition, even when the answers were obtained, they were very verbose, and were not very intuitive or crisp. The participants felt that there was a huge cognitive load in using search engines to obtain relevant information. When they used our search interface, they were happy that the information could be obtained very quickly, and without taxing them much.

We are planning for a more elaborate user study to collect more quantitative feedback about our system.

Based on our initial prototypes, we believe that there exists a viable business model that can help apply semantic web techniques to real life scenario.

**FUTURE WORK**

There are many possible research activities around this work. Several extensions are possible and these include:

- Interactive creation of responses**:** Stored information is not always sufficient to answer end-user queries; interaction with information suppliers, typically vendors of products and services, may be necessary to answer some queries. One way of extending the system to meet this need economically is to have the end user to indicate what he needs, say the best price for a handycam identified by its maker and model number. The system could send a query over the Web or the cellular network to each vendor dealing with this kind of product to get a response in a specified form. The system could then present these in a suitable order to the end user. Since it may not be possible to get responses in real time, the user could be given a response, say, 10 or 15 minutes later over email accompanied by an SMS alert.

- Automatic generation of user interface screens based on an ontology: Currently, we are creating a user interface for every ontology and using this interface to provide inputs. However, if we can automatically generate a user interface for any ontology, then it will be possible to easily extend this system to many more domains/businesses.

- Query Distribution and Response reassembly: When a query is received by the query engine, it is likely that the user wants information that may be distributed over multiple information stores. In these scenarios, we might need to distribute the semantic query to a number of appropriate servers. When responses are received from each of these servers, we need to reassemble the responses and send a single response back to the user.

- Automatic import of business data from various web sites: It is likely that the business establishments have data available in their respective web sites, and they would want this to be reused. In order to do this, we need to carry out conversion of selected database content to RDF triple form. Customers could follow specified standards to create triple stores that can be periodically harvested.

- Reasoning: Our work can be extended to include deeper reasoning which can help the user to increase the chances of getting a good response from the system going beyond simple and direct answers. Currently, we only use shallow reasoning to increase the recall in our system.

- Automatic Creation of Ontology: We are looking into the possibility of creating an ontology using already existing data on the web, using semi-automatic conversion techniques. This would require fairly sophisticated natural language processing. This would enable service providers to import data from various businesses that already have a presence on the web. Techniques would need to be developed to let customers examine and approve converted data.

## CONCLUSION

The case made above is that it is possible and highly desirable to create systems which will enable the creation of structured information by interested parties in the spirit of the Web 2.0. We have provided for access to information from mobile phones and other mobile devices as well as PCs on the Web, and implemented very easy to use human computer interfaces. These systems vastly reduce the need for human intervention, and as a result potentially reduce costs and improve regular updating of information. We have indicated the arguments in favor of using semantic web technology to implement these systems.

Using vendor independent protocols and standards, semantic web technology provides for federating of information from multiple servers to provide a greater variety of information to the end user, and to create mashups which put together in a coherent manner information coming from multiple semantic end points.